\documentclass[aps,prd,natbib,floatfix,noshowkeys,epsfig,graphics]{revtex4}%
\usepackage{graphicx}
\usepackage{amsmath}
\usepackage{amsfonts}
\usepackage{amssymb}
\usepackage{kotex}
\usepackage{xcolor}
\usepackage{ulem}
\usepackage{comment}
\usepackage{multirow}
\usepackage[toc,page]
{appendix}

\newcommand{\newc}{\newcommand}

\newc{\bs}    {\section}
\newc{\no}    {\\ \nonumber}

\newc{\st}    {\stackrel}

\begin{document}

\title{Dynamical friction for circular orbits in self-interacting ultralight dark matter and Fornax globular clusters}

\author{Hyeonmo Koo$^{ \tt a}$, Jae-Weon Lee$^{ \tt b}$ }
\email{mike1919@uos.ac.kr, scikid@jwu.ac.kr}


\affiliation{ \centerline{\sl a) Physics Department,
University of Seoul, Seoul 02504, Korea}
 \centerline{\sl b) 
 Department of Electrical and Electronic Engineering, Jungwon University,
 85 Munmuro, Goesan, Chungbuk 28024, Korea}
 }

\date{\today}
\begin{abstract}
We investigate the impact of repulsive self-interaction in ultralight dark matter (ULDM) on dynamical friction in circular orbits in ULDM halos and its implications for the Fornax dwarf spheroidal (dSph) galaxy's globular clusters. 
Using the Gross-Pitaevskii-Poisson equations, we derive the dynamical friction force considering soliton density profiles for both non-interacting and strongly self-interacting ULDM. 
Our results show that self-interactions reduce the dynamical friction effect further than both the non-interacting ULDM and standard cold dark matter models.
Furthermore, we derive the low Mach number approximation to simplify the analysis in the subsonic motion, where the tangential component of dynamical friction dominates.
Applying these findings to the Fornax dSph, we calculate the infall timescales of globular clusters, demonstrating that strong self-interaction can address the timing problem more effectively. 
We constrain the parameter space for ULDM particle mass and self-coupling constant, which are consistent with other constraints from astronomical and cosmological observations.
\end{abstract}

\maketitle

\section{Introduction}\label{sec1}
Ultralight dark matter (ULDM), also known as fuzzy dark matter, scalar field dark matter or the ultralight axion~\cite{Baldeschi:1983mq,Sin:1992bg,Lee:1995af,Matos:1998vk,Hu:2000ke,Boehmer:2007um,Matos:2023usa}, has recently emerged as an alternative to cold dark matter (CDM). 
While the CDM model serves as the fundamental basis of the standard cosmological model, it faces  challenges in accounting for galactic-scale structures, notably the core-cusp discrepancy and the missing satellite problem \cite{Hu:2000ke,Salucci:2002nc,Park:2022lel,Koo:2023gfm}.
In this model, to address small-scale issues, dark matter particles are assumed to have an extraordinarily small mass $m_\phi$, typically of the order of $10^{-22} ~\mathrm{eV}$, and 
ULDM is often modeled as a classical scalar field, where quantum corrections are negligible \cite{Hui:2016ltb}.

The cosmological constraints on ULDM are highly restricted in the case of no self-interaction (often called the fuzzy dark matter (FDM)), leading to a narrow allowed parameter space~\cite{Ferreira:2020fam}, which faces some observational challenges, particularly in explaining features of the Lyman-$\alpha$ forest~\cite{Irsic:2017yje,Armengaud:2017nkf}. 
Cosmic microwave background (CMB) anisotropies and large-scale structure measurements disfavor masses below $10^{-23} ~\mathrm{eV}$~\cite{Bozek_2015}, and weak gravitational lensing constraints from the Dark Energy Survey (DES) together with Planck CMB results likewise require $m_\phi \gtrsim 10^{-23} ~\mathrm{eV}$~\cite{Dentler_2022}.
On the other hand, Lyman-$\alpha$ forest data combined with SDSS and high-resolution spectra disfavor $10^{-22} ~\mathrm{eV} \lesssim m_\phi < (2$-$3)\times 10^{-21} ~\mathrm{eV}$ at 95\% C.L.~\cite{Irsic:2017yje, Armengaud:2017nkf}. 
Independent probes using subhalo mass functions from strong lensing and stellar streams also suggest $m_\phi \gtrsim 2.1\times 10^{-21}\, \mathrm{eV}$ \cite{Schutz:2020jox}, while recent cosmological zoom-in simulations (COZMIC) that directly model Milky Way (MW) satellite populations further strengthen this limit to $m_\phi \gtrsim 1.4\times 10^{-20}\,\mathrm{eV}$~\cite{Nadler:2025fcv}.
Complementary analyses based on stellar kinematics of dwarf galaxies also provide stringent limits on the ULDM particle mass, from ultra-faint systems such as Segue 1 and Segue 2 yielding $m_\phi \gtrsim 3\times 10^{-19}~\mathrm{eV}$ \cite{Dalal:2022rmp} to numerical simulations of Fornax-like dwarfs indicating tensions for $m_\phi \lesssim 5\times 10^{-21}~\mathrm{eV}$ \cite{Teodori:2025rul}.
Galactic rotation curve analyses using the SPARC database further disfavor the soliton–halo relation of ULDM simulations across $10^{-24}~\mathrm{eV} \lesssim m_\phi \lesssim 10^{-20}~\mathrm{eV}$ \cite{Bar:2021kti}.
The current landscape of constraints has been comprehensively reviewed in~\cite{Eberhardt:2025caq}.

Incorporating self-interaction in ULDM~\cite{Lee:1995af,Boehmer:2007um,Chavanis:2011zi} offers a promising avenue to alleviate these tensions in mass and expand the model's compatibility with observations~\cite{Dave:2023wjq,Li:2013nal,Lee:2024pxl}.
Self-interactions in ULDM introduce a repulsive force that significantly modifies soliton density profiles, enabling a broader range of dynamical behaviors on galactic scales even for dimensionless self-coupling constants as small as $\lambda\sim 10^{-90}$ \cite{Glennon:2022huu}.
Furthermore, self-interacting ULDM is recently  proposed as a mechanism for generating neutrino mass and electroweak scales~\cite{Lee:2024rdc}, and as a solution to the Hubble tension~\cite{Lee:2025roi}.

These interactions profoundly affect key astrophysical processes, particularly dynamical friction (DF), which governs the orbital evolution of massive objects within dark matter halos. 
Understanding DF of dark matter is crucial for addressing unresolved challenges, such as the timing problem of the Fornax dwarf spheroidal galaxy's globular clusters.
The timing problem \cite{Fornax1Gyr,Tremaine_nuclei1,Tremaine_nuclei2} arises from the observed longevity of these clusters in peripheral orbits, whereas classical DF in CDM predicts that they should have decayed into the galaxy's center within a fraction of their current lifetimes. 
While FDM has been proposed as a potential solution due to its weaker DF force compared to CDM \cite{Lancaster:2019mde, Hui:2016ltb}, the inclusion of self-interaction may further extend the infall timescales \cite{Glennon:2023gfm, Hartman:2020fbg}, offering a more comprehensive explanation.

In this work, we explore the effects of self-interaction in ULDM on DF for objects in circular orbits within ULDM halos, focusing on its implications for the Fornax dSph galaxy. We derive the DF force using a hydrodynamic formulation of the Gross-Pitaevskii-Poisson equations and compare results between non-interacting and self-interacting regimes. The results allow us to constrain the particle mass and self-coupling constant, through comparison with observed properties of Fornax globular clusters (GCs). 

In Section II, we present ULDM soliton profiles for non-interacting and strongly self-interacting regimes. 
In Section III, we review the derivation of DF force~\cite{Hui:2016ltb,Berezhiani:2023vlo} using the GPP equations and study its dependence on self-interactions and the low-Mach number approximation.
In Section IV, we apply the DF framework to the Fornax dwarf spheroidal galaxy, calculating infall timescales for GCs and constraining ULDM parameters.
In Section V, we summarize the findings and discuss the implications for dark matter physics and astrophysical observations.

\section{ultralight dark matter models and their soliton profiles}\label{sec2}

In this paper ULDM is modeled as a real scalar field $\phi$ with the following effective action~\cite{Lee:1995af, Chavanis:2011zi}:
\begin{equation}
S=\int d^4 x \sqrt{-g} \left[ \frac{R}{16\pi G} - \frac{1}{2} g^{\mu\nu}\left(\partial_\mu\phi\right) \left(\partial_\nu\phi\right) - V(\phi) \right]
\label{Effective_Action}
\end{equation}
where the potential, including the mass term and a repulsive self-interaction, is expressed as:
\begin{equation}
V(\phi)=\frac{m_\phi^2 c^2}{2\hbar^2} \phi^2 + \frac{\lambda}{4\hbar c}\phi^4,
\label{phi4-potential}
\end{equation}
where $m_\phi$ is the particle mass and $\lambda$ is the dimensionless self-coupling constant.
The evolution of the scalar field is described by the Klein-Gordon equation
\begin{equation}
\square \phi - \frac{dV}{d\phi} = 0,
\label{Klein-Gordon}
\end{equation}
where $\square$ is the d'Alembertian, and the gravitational potential $\Phi_\mathrm{U}$ of $\phi$ is encoded in the metric.
In the galactic-scale regime, the relativistic mode is factored out from $\phi$ by introducing a slowly-varying complex scalar field $\psi$~\cite{Hui:2016ltb, Hui:2021tkt}, given by
\begin{equation}
\phi({\bf x},t)=\frac{\hbar}{\sqrt{2 m_\phi}}[e^{-im_\phi c^2 t/\hbar}\psi({\bf x},t) + e^{+im_\phi c^2 t/\hbar}\psi^*({\bf x},t)].
\label{nonrelativistic_decomposition}
\end{equation}
Including perturbation via the potential $\Phi_\mathrm{P}$, 
which represents the gravitational potential of a compact object such as a black hole or a globular cluster inside DM,
the equations of motion take the form of Gross-Pitaevskii-Poisson (GPP) system~\cite{Lee:1995af, Boehmer:2007um, Chavanis:2011zi, Rindler-Daller:2013zxa}:
\begin{align}
i\hbar \frac{\partial}{\partial t} \psi({\bf x},t ) & = -\frac{\hbar^2}{2m_\phi}\nabla^2 \psi({\bf x},t )  +m_\phi [\Phi_\mathrm{U}({\bf x},t )+\Phi_\mathrm{P}({\bf x},t )] \psi({\bf x},t )+\frac{\hbar^3 \lambda}{2m_\phi^2 c}|\psi|^2({\bf x},t ) \psi({\bf x},t ),
\label{GPP_GPeq}
\\
 \nabla^2 \Phi_\mathrm{U}({\bf x},t )&=4\pi G m_\phi |\psi|^2({\bf x},t ),
\label{GPP_Peq}
\end{align}
where $\psi({\bf x}, t)$ represents a classical field configuration with mass density $\rho=m_\phi |\psi|^2$.
A time-independent, spherically symmetric, ground-state solution of the GPP equations has a ``Soliton'' structure, balanced by gravitational collapse, quantum pressure, and barotropic pressure from ULDM self-interaction.
Obtaining an exact soliton solution of the GPP equations analytically is challenging.
Instead, several models are used to describe the spatial density profile of the soliton in various limiting cases.

First, in the FDM regime ($\lambda = 0$), the system of equations \ref{GPP_GPeq} and \ref{GPP_Peq} reduces to the simpler form, the Schrödinger-Poisson equations~\cite{Hu:2000ke, Hui:2021tkt, Hui:2016ltb}.
In this regime, the balance between gravitational collapse and quantum pressure leads to the stability of a soliton.
The ground-state solution is approximated to the empirical density profile~\cite{Schive:2014dra} 
\begin{equation}
\rho(r)=\frac{\rho_c}{[1+\alpha (r/r_c)^2]^8},
\label{Empirical_Profile}
\end{equation}
where $\rho_c$ is the central density, $r_c$ is the core radius defined as $\rho(r_c)=\rho_c/2$, and $\alpha=2^{1/8}-1 \simeq 0.091$.
For a given soliton mass $M_\mathrm{sol}$, estimates of $\rho_c$ and $r_c$ are given by~\cite{Schive:2014dra, Indjin:2023hno,Hui:2016ltb}
\begin{equation}
\rho_c = 0.019 \ M_\odot / \mathrm{pc}^3  \left(\frac{10^{-22}~\mathrm{eV}}{m_\phi}\right)^2 \left(\frac{1\mathrm{kpc}}{r_c}\right)^4
\ \ \ , \ \ \
r_c=0.2278 ~\mathrm{kpc} \left(\frac{10^9 M_\odot}{M_\mathrm{sol}} \right) \left(\frac{10^{-22}~\mathrm{eV}}{m_\phi}\right)^2.
\label{param_emp}
\end{equation}
Since the empirical profile provides a reliable approximation to the ground-state of ULDM in the non-interacting regime, these estimates allow one to construct a density profile for a given combination of soliton mass and ULDM particle mass $(M_\mathrm{sol},m_\phi)$, reflecting the characteristic scales emerging from the classical field approximation governed by the Schrödinger-Poisson equations \cite{Lee:2023krm}.

Second, in the strongly self-interacting regime ($\lambda\gg0$), we consider the Thomas-Fermi limit for the soliton profile, which corresponds to neglecting the quantum pressure.
In this regime, gravitational collapse is balanced solely by barotropic pressure.
This approximation is also consistent with neglecting terms of $\sim\nabla^2\psi$ and $\sim\frac{\partial\psi}{\partial t}$ in Eq. \ref{GPP_GPeq}, and the exact ground-state solution is given by~\cite{Lee:1995af, Boehmer:2007um, Chavanis:2011zi}
\begin{equation}
\rho(r)=\rho_c\frac{\sin(\pi r/R_\mathrm{TF})}{(\pi r/R_\mathrm{TF})},
\label{Thomas-Fermi_Limit}
\end{equation}
where $\rho_c$ and soliton size $R_\mathrm{TF}$ are
\begin{equation}
\rho_c = \frac{\pi M_\mathrm{sol}}{4R_\mathrm{TF}^3} 
\ \ \ , \ \ \
R_\mathrm{TF}=\sqrt{\frac{\pi \hbar^3 \lambda}{8Gm_\phi^4 c}}.
\label{param_TFlim}
\end{equation}
Note that $R_\mathrm{TF}$ is a constant for
given $\lambda$ and $m_\phi$, while the size of the soliton in the FDM model is inversely proportional to the soliton mass. 
The validity of the Thomas-Fermi limit is further examined in Appendix \ref{App_Validity} through dimensional analysis based on the characteristic length scale of the soliton.
In particular, this approximation breaks down for excessively small self-interaction, where quantum pressure can no longer be neglected.

\section{Dynamical friction}\label{sec3}

We consider a small object orbiting at a constant radius $r_0$ with angular velocity $\Omega$ as a perturbing source in the ULDM medium.
In other words, the position of the object is given by ${\bf x}_\mathrm{P}(t)=r_0 (\hat{\bf x} \cos \Omega t  + \hat{\bf y} \sin \Omega t )$ for time $t$, with rotational speed $v_0\equiv \Omega r_0$.
In this section, we calculate the dynamical friction (DF) force, ${\bf F}_\mathrm{DF}$, acting on the object, which arises from the ULDM over-density induced by its gravitational potential $\Phi_\mathrm{P}$.
From the line-of-sight velocity
of the GCs relative to Fornax itself
~\cite{Cole:2012ns}
one can estimate
$v_0=O(10) ~\mathrm{km/s}$
for the Fornax GCs.

\subsection{Ultralight dark matter over-density from Madelung formalism}\label{sec3.A}
The Madelung formalism describes the ULDM system by decomposing the wavefunction as $\psi({\bf x},t)=\sqrt{\frac{\rho({\bf x},t)}{m_\phi}} e^{i\theta({\bf x},t)}$ with the velocity field ${\bf v}({\bf x},t)=\frac{\hbar}{m_\phi}\nabla\theta({\bf x},t)$~\cite{Hui:2021tkt, Hui:2016ltb}.
This decomposition leads to the continuity and Euler equations, which are fundamental in fluid dynamics:
\begin{align}
\frac{\partial \rho}{\partial t}+ \nabla \cdot (\rho {\bf v})& = 0,
\label{Continuity_eq}
\\
\frac{\partial {\bf v}}{\partial t} + ({\bf v} \cdot \nabla) {\bf v} &=- \nabla Q-\frac{1}{\rho}\nabla P  - \nabla (\Phi_\mathrm{U}+\Phi_\mathrm{P}),
\label{Euler_eq}
\end{align}
where $Q=-\frac{\hbar^2}{2m_\phi^2} \frac{\nabla^2 \sqrt{\rho}}{\sqrt{\rho}}$ is commonly referred to as the quantum pressure, and $P=\frac{\lambda \hbar^3}{4m_\phi^4 c}\rho^2$ is the barotropic pressure arising from ULDM self-interaction.
We consider the fractional over-density of ULDM fluid given by
\begin{equation}
\alpha({\bf x},t)\equiv \frac{\rho({\bf x},t)}{\bar{\rho}}-1 ,
\label{Linear_Density_Perturbation}
\end{equation}
where $\bar{\rho}$ is the background density.
We regard the soliton central density, $\rho_c$, as the background density, which is approximately constant in our work.
The adiabatic sound speed $c_s$ for barotropic pressure in the ULDM background is given by
\begin{equation}
c_s^2=\left(\frac{\delta P}{\delta \rho}\right) _{\rho_c}=\frac{\lambda \hbar^3}{2m_\phi^4 c}\rho_c.
\label{sound_speed}
\end{equation}
This suggests that in the strongly self-interacting regime, the ULDM behaves as a barotropic fluid characterized by a sound speed $c_s$, often referred to as the sound regime~\cite{DFrot_Gas, Berezhiani:2023vlo}.
By combining equations \ref{Continuity_eq} $\sim$  \ref{sound_speed}, the evolution equation of the over-density is then
\begin{equation}
\frac{\partial^2\alpha}{\partial t^2} - c_s^2 \nabla^2 \alpha + \frac{\hbar^2}{4m_\phi^2}\nabla^4 \alpha = \nabla^2 \Phi_\mathrm{P}.
\label{evolution_density_perturbation}
\end{equation}
We neglect the effect of ULDM self-gravity $\Phi_\mathrm{U}$ for describing the evolution of the over-density.
While self-gravity is effective at low-momentum modes in the dispersion relation of Eq. \ref{evolution_density_perturbation}~\cite{Berezhiani:2023vlo}, which is relevant for soliton structure formation, it does not significantly affect the gravitational wake responsible for the DF of point-like particles orbiting ULDM, making its contribution negligible in this context~\cite{Hui:2016ltb}.
This is because the density perturbations responsible for DF are dominated by high-momentum modes, where the self-gravity term becomes subdominant.
Therefore, Eq. \ref{evolution_density_perturbation} can be solved using the Green function method, yielding
\begin{equation}
\alpha({\bf x},t)=\int d^3 {\bf x}' dt' G({\bf x}-{\bf x}', t-t') \nabla^2 \Phi_\mathrm{P}({\bf x}',t').
\label{overdensity_GreenFunction}
\end{equation}
The Green function $G$ can be expressed as the Fourier transform
\begin{equation}
G({\bf R},\tau)=\int \frac{d^3 {\bf k}}{(2\pi)^3}\frac{d\omega}{2\pi} \frac{e^{i({\bf k}\cdot{\bf R} - \omega\tau)}}{\frac{\hbar^2}{4m_\phi^2}k^4 + c_s^2 k^2 - (\omega+i\epsilon)^2}
\label{GreenFunction}
\end{equation}
with the condition $\epsilon>0$ enforcing causality of the dynamics.

\subsection{Calculation of dynamical friction force}\label{sec3.B}
In this subsection, we briefly review the calculation of the DF force by ULDM acting on the circular orbit~\cite{Berezhiani:2023vlo}.
This formulation is applicable to various astrophysical phenomena that assume circular orbits.
Extension of this framework to elliptical orbits have been explored in the context of gaseous media, though not yet for ULDM~\cite{DFrot_Elliptic}.

The DF force induced by the over-density $\alpha({\bf x},t)$ is expressed as
\begin{equation}
{\bf F}_\mathrm{DF}(t)=\rho_c\int d^3 {\bf x}(\nabla\Phi_\mathrm{P})\alpha({\bf x},t).
\label{DF_definition}
\end{equation}
We consider a point-like object of mass $M_\mathrm{P}$: $\Phi_\mathrm{P}=-GM_\mathrm{P} \frac{1}{|{\bf x}-{\bf x}_\mathrm{P}(t)|}$, 
 as outlined in~\cite{DFrot_Gas, DFrot_FDM}.
We combine equations \ref{overdensity_GreenFunction} $\sim$ \ref{DF_definition} with defining the variable ${\bf u}\equiv {\bf x}-{\bf x}_\mathrm{P}(t)$ to use the integral $\int d^3{\bf u} \frac{\bf u}{u^3}e^{i \bf k \cdot u}=4\pi i \frac{\bf k}{k^2}$, and split the integral over $\bf k$ into radial ($k$) and angular ($\hat{\bf k}$) parts, respectively, to simplify the angular dependence of the gravitational potential's Fourier transform.
Then, the DF force is written as
\begin{align}
{\bf F}_\mathrm{DF}(t)&=(4\pi G M_\mathrm{P})^2\rho_c \int d\tau \frac{d\omega}{2\pi} e^{-i \omega\tau} \int \frac{d^3 {\bf k}}{(2\pi)^3} \frac{i\bf k}{k^2} \frac{e^{i{\bf k}\cdot[{\bf x}_\mathrm{P}(t)-{\bf x}_\mathrm{P}(t-\tau)]}}{\frac{\hbar^2}{4m_\phi^2}k^4 + c_s^2 k^2 - (\omega+i\epsilon)^2}\label{DF_integral} \\
&=4\pi(G M_\mathrm{P})^2\rho_c \int d\tau \frac{d\omega}{2\pi} e^{-i \omega\tau} \int_0^\infty 
\frac{kdk}{\frac{\hbar^2}{4m_\phi^2}k^4 + c_s^2 k^2 - (\omega+i\epsilon)^2} \left\{\int \frac{d^2 \hat{\bf k}}{2\pi^2} i\hat{\bf k}e^{i{\bf k}\cdot[{\bf x}_\mathrm{P}(t)-{\bf x}_\mathrm{P}(t-\tau)]}\right\} .
\label{DF_integral_split}
\end{align}
An essential difference lies in the treatment of ${\bf x}_\mathrm{P}$ in the angular integral inside \{\}, as previous studies~\cite{Lancaster:2019mde, Hui:2016ltb} focus on linear motion, whereas we calculate the DF for circular motion.
Details of the calculation of the angular integral, using Rayleigh expansion of exponentials, are explained in~\cite{Berezhiani:2023vlo}.
The resulting DF force can be expressed as
\begin{equation}
{\bf F}_\mathrm{DF}(t)=-4\pi \rho_c\left(\frac{GM_\mathrm{P}}{v_0}\right)^2\left[ 
\Re(I)\hat{\bf r}(t) + \Im(I) \hat{\bf \phi}(t)\right].
\label{Result_F}
\end{equation}
The dimensionless factor $I$ separately contributes to the DF force.
First, the real part ($\Re(I)$) represents the DF coefficient for the radial direction ($\hat{\bf r}$), which points toward the center of the orbit.
Second, the imaginary part ($\Im(I)$) represents for the tangential direction ($\hat{\bf \phi}$) of the orbit.
Also, $I$ is expressed as a sum over angular multipoles $(l,m)$:
\begin{equation}
I=\sum_{\ell=1}^{\ell_\mathrm{max}} \sum_{m=-\ell}^{\ell-2} (-1)^{m} \frac{(\ell-m)!}{(\ell-m-2)!} \frac{S_{\ell,\ell-1}^m - S_{\ell,\ell-1}^{-m-1}}{\Gamma\left(\frac{1-\ell-m}{2}\right)  \Gamma\left(\frac{2+\ell-m}{2}\right) \Gamma\left(\frac{3-\ell+m}{2}\right) \Gamma\left(\frac{2+\ell+m}{2}\right)}
\label{Result_I}
\end{equation}
with the Gamma function $\Gamma(z)$ and 
\begin{equation}
S_{\ell_1,\ell_2}^m = v_0^2 \int_0^{\infty} dk \frac{k j_{\ell_1} (kr_0) j_{\ell_2} (kr_0)}{\frac{\hbar^2}{4m_\phi^2}k^4 + c_s^2 k^2 - (m\Omega+i\epsilon)^2}.
\label{integral_S}
\end{equation}
The quantity $S_{\ell_1,\ell_2}^{m}$ represents the ``Scattering amplitude'' of the radial wave, associated with the $(\ell_1,\ell_2,m)$-th component of the partial wave.
Detail of its calculation, applying Cauchy's integral formula, is discussed in Appendix B of~\cite{Berezhiani:2023vlo}.
During the evaluation, two dimensionless variables are required to describe the dynamics.
One is the Mach number of moving object inside ULDM fluid, defined as
\begin{equation}
\mathcal{M}\equiv \frac{v_0}{c_s},
\label{Mach_Number}
\end{equation}
which characterizes whether the motion is subsonic ($\mathcal{M}<1$) or supersonic ($\mathcal{M}>1$).
Another variable is 
\begin{equation}
\ell_q \equiv \frac{m_\phi c_s r_0}{\hbar\mathcal{M}},
\label{l_q}
\end{equation}
which represents the importance of self-interaction relative to the quantum pressure effects~\cite{Berezhiani:2023vlo}.
The four poles in the contour integral of Eq. \ref{integral_S} naturally define two quantities: $f_m^\pm\equiv \sqrt{2\sqrt{1+\frac{m^2}{\ell_q^2}}\pm 2}$.
In the sound regime ($\ell_q \gg 1$), self-interaction dominates and perturbations propagate as sound wave, with $f_m^+\rightarrow 2$ and $f_m^-\rightarrow \frac{m}{\ell_q}$.
Conversely, in the quantum regime ($\ell_q\ll 1$), the dynamics are governed by quantum pressure, leading to suppressed density perturbations, with $f_m^\pm \rightarrow \sqrt{\frac{2m}{\ell_q}}$.
Finally, the result is~\cite{Berezhiani:2023vlo, Gorkavenko:2024ocl}
\begin{equation}
S_{\ell,\ell-1}^m=\mathcal{M}^2 \times \begin{cases}
\frac{\pi i}{2\sqrt{1+m^2/\ell_q^2}} \left[j_{\ell}(\ell_q \mathcal{M}f_m^-) h_{\ell-1}^{(1)}(\ell_q \mathcal{M}f_m^-) - j_{\ell}(i\ell_q \mathcal{M}f_m^+) h_{\ell-1}^{(1)}(i\ell_q \mathcal{M}f_m^+) \right] & (m>0)\\
\frac{-\pi i}{2\sqrt{1+m^2/\ell_q^2}} \left[j_{\ell}(\ell_q \mathcal{M}f_m^-) h_{\ell-1}^{(2)}(\ell_q \mathcal{M}f_m^-) + j_{\ell}(i\ell_q \mathcal{M}f_m^+) h_{\ell-1}^{(1)}(i\ell_q \mathcal{M}f_m^+) \right] & (m<0)\\
\frac{\pi}{2}\left[ \frac{1}{4\ell^2-1}-i j_{\ell}(2i\ell_q \mathcal{M}) h_{\ell-1}^{(1)}(2i\ell_q \mathcal{M}) \right] & (m=0)
\end{cases}
\label{Result_S}
\end{equation}
This expression captures the scattering behavior of density perturbations induced by the orbiting object, which determines the strength and directionality of the resulting DF force.
Note that $S_{\ell,\ell-1}^{m=0}$ is real and does not contribute to $\Im(I)$.
Results of $\Re(I)$ and $\Im(I)$ for $\ell_q \mathcal{M}=3$ and different maximum multipoles are plotted in the Fig. 3 of~\cite{Berezhiani:2023vlo}.
This shows that the radial component of DF force is suppressed for subsonic motion rather than the tangential component.

\subsection{Low-Mach number approximation of dynamical friction coefficient}\label{sec3.C}

Both the calculations of the multipole expansion in Eq. \ref{Result_I} and the spherical Bessel functions in Eq. \ref{Result_S} are computationally demanding.
To address this computational complexity, we adopt a leading-order approximation in the low-Mach number regime.
This approach simplifies the analysis while capturing the dominant contributions to the DF force in $\mathcal{M}<1$.
Our study naturally falls within the subsonic motion.
From Eq. (\ref{sound_speed}) one can obtain
\begin{equation}
   \mathcal{M}= \frac{v_0}{\sqrt{\frac{\lambda \hbar^3}{2m_\phi^4 c}{\rho_c}}},
\end{equation}
which gives 
$\mathcal{M}=0.0087$
for typical values
$v_0\simeq 10~\mathrm{km/s}$ and
$\rho_c\simeq 0.1~M_\odot/\mathrm{pc}^3$ for the 
globular clusters
~\cite{Read:2018fxs}.

\begin{figure}[!ht]
\centering
\includegraphics[width=0.5\textwidth]{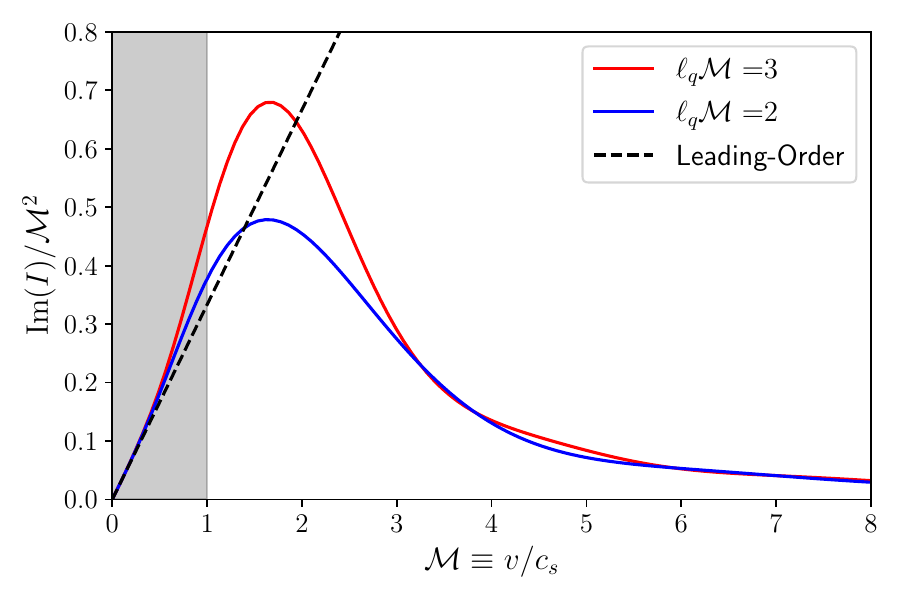}
 \caption{
 Tangential DF coefficients for $\ell_q \mathcal{M}=3$ (red), 2 (blue), and their leading-order term (black dashed).
 For subsonic motion (shaded region), this term could mainly illustrates the behavior of full-order calculation, better for lower $\ell_q \mathcal{M}$.
 }
 \label{DFcoeff}
\end{figure}

For the low-Mach number approximation, two notable characteristics emerge, particularly regarding the dominance of the tangential DF force.
First, the leading-order term of $\Im(I)$ appears only for $\ell=1$, making the multipole $(\ell,m)=(1,-1)$ the sole significant contribution.
Second, the asymptotic behavior of $\ell_q$ and $f_{-1}^\pm$ leads to $\ell_q \mathcal{M}f_{-1}^+ \simeq \frac{2m_\phi c_s r_0}{\hbar}$ and $\ell_q \mathcal{M}f_{-1}^- \simeq \mathcal{M}$.
The detailed calculation of the leading-order term of $S_{\ell,\ell-1}^m$ is deferred to Appendix \ref{App_sphericalbessel} for clarity and completeness.
The result of the leading-order term for the tangential DF coefficient $\Im(I)$ is
\begin{equation}
\Im(I)_{\ell=1}=\frac{1}{3}\mathcal{M}^3 + O(\mathcal{M}^5).
\label{ImI_l=1}
\end{equation}
This expression provides the DF coefficients under the low-Mach number approximation, where we retain only the leading-order terms in $\mathcal{M}$.
We find that the lowest-order term of the radial DF coefficient for $\ell_q \gg 1$ is of order $\Re(I)\sim O(\mathcal{M}^4)$ and occurs for multiple $\ell$ values, not just for $\ell=1$.
This indicates that the leading-order contribution of $\Im(I)$ relative to $\Re(I)$ implies the tangential component to dominate the DF in subsonic motion, making it more significant than the radial component.
Figure \ref{DFcoeff} illustrates $\Im(I)$ as a function of $\mathcal{M}$ for $\ell_q \mathcal{M}=2,~3$, along with their leading-order approximations.
For fixed orbital radius and angular velocity, $\ell_q \mathcal{M}=\frac{m_\phi c_s r_0}{\hbar}$ serves as a dimensionless measure of self-coupling constant.
Setting $\ell_\mathrm{max}=20$ ensures sufficient convergence of the DF coefficients, numerically verified in~\cite{Berezhiani:2023vlo}.
In the subsonic motion ($\mathcal{M}<1$), the full-order DF coefficients closely match their leading-order approximations and increase steadily with $\mathcal{M}$.
However, the DF coefficients decrease after reaching a maximum at $\mathcal{M}\lesssim 2$, indicating the breakdown of the leading-order approximations.
This approximation can be applied to the five GCs orbiting the Fornax dSph, which will be discussed in the next section.

\subsection{Dynamical friction coefficient for non-interacting limit}\label{sec3.D}

The non-interacting limit corresponds to the regime $\ell_q \ll 1$, where quantum pressure dominates as self-interactions are negligible.
In this regime, the system demonstrates the characteristic suppression of DF observed in FDM due to its wave-like nature,
with $f_m^\pm \rightarrow \sqrt{\frac{2m}{\ell_q}}$ where $S_{\ell,\ell-1}^m$ is given by~\cite{DFrot_FDM}:
\begin{equation}
S_{\ell,\ell-1}^{m\ne 0} =\frac{i\pi R_\Omega}{4m}\left[
j_{\ell}(\sqrt{mR_\Omega}) h_{\ell-1}^{(1)}(\sqrt{mR_\Omega}) - j_{\ell}(i\sqrt{mR_\Omega}) h_{\ell-1}^{(1)}(i\sqrt{mR_\Omega})
\right].
\label{Result_S_FDM}
\end{equation}
Here, $R_\Omega$ represents the ratio of the orbit size ($2r_0$) to the de Broglie wavelength ($\lambdabar_\Omega \equiv \frac{\hbar}{m_\phi v_0}$): $R_\Omega \equiv \frac{2m_\phi v_0 r_0 }{\hbar}$, which can be interpreted as a characteristic angular momentum scale that governs the system's quantum effects, already formulated in~\cite{DFrot_FDM}.
In this work, we focus on the imaginary part of $S_{\ell,\ell-1}^m$, which remains well-defined and plays a key role in the DF calculation.

As in the previous subsection, the asymptotic expansion in the low-$R_\Omega$ limit is crucial for understanding the leading-order contribution.
This describes the low-$R_\Omega$ approximation in FDM, where the de Broglie wavelength is comparable to the orbital scale.
When applying Eq. \ref{Result_I} to Eq. \ref{Result_S_FDM}, the lowest-order term for $R_\Omega$ arises only for $\ell=1$, as higher-order terms are suppressed.
The dominant term for $\Im(I)$ is then
\begin{equation}
\Im(I)_{\ell=1}=\frac{1}{6}R_\Omega^{3/2} + O(R_\Omega^{5/2}).
\label{ImI_l=1_FDM}
\end{equation}
This result represents the DF coefficient obtained from the low-$R_\Omega$ approximation.
We find that the leading-order approximation gives $\Re(I)\sim O(R_\Omega^2)$, appearing not only for $\ell=1$ but also for higher mutlipoles.
This indicates that the radial DF component is subdominant, and thus the overall DF is governed primarily by its tangential component.
We also apply this to the five GCs orbiting the Fornax dSph in the next section.

\section{Application to Fornax globular clusters}\label{sec4}
The Fornax dwarf spheroidal (dSph) galaxy, where the halo mass is estimated as $M_\mathrm{Fornax}=1.42\times 10^8 M_\odot$, is one of the most massive and low-luminosity satellites orbiting the MW, and it hosts six globular clusters (GC).
Its significant DM dominance and negligible tidal disruption make it an ideal target for studying DM dynamics~\cite{Coleman:2008kk, Cole:2012ns}.
Observational data confirm that the GCs have maintained stable orbits far from the galactic center for $\sim 10~\mathrm{Gyr}$~\cite{DES:2018jtu, delPino:2013nfa}, which are referred to as their lifetimes.
However, the classical DF model within the CDM framework~\cite{Chandrasekhar1943} predicts that these GCs are expected to spiral into the Fornax nucleus within decay timescales of $\sim 1~\mathrm{Gyr}$~\cite{Fornax1Gyr}, significantly shorter than their estimated lifetimes.
This is commonly referred to as the timing problem of Fornax GCs~\cite{Tremaine_nuclei1,Tremaine_nuclei2}.
Previous studies have shown that the FDM model can solve this problem due to a weaker DF force compared to the CDM model~\cite{Lancaster:2019mde, DFrot_FDM}.
This suppression arises from the wave nature of FDM, which prevents the efficient formation of gravitational wakes typically responsible for DF force.
The addition of repulsive self-interaction in ULDM is expected to enhance the effective sound speed, thereby further reducing DF force and extending the infall timescales of GCs.

Numerical simulations show that MW-like DM halos consist of a central soliton core surrounded by an outer envelope with an approximately $r^{-3}$ density profile, resembling the NFW profile~\cite{Schive:2014hza, Schwabe:2016rze, Mocz:2017wlg}.
These studies also suggest that, for halos with masses on the order of $\sim 10^8 M_\odot$, the soliton core can dominate the overall mass distribution of the halo.
However, this does not imply that the soliton core mass ($M_\mathrm{sol}$) is identical to the total halo mass ($M_h$).
Mocz et al. \cite{Mocz:2017wlg} demonstrated through numerical simulations that the core-halo mass relation in FDM satisfies
\begin{equation}
M_{\mathrm{sol}}=1.31\times 10^{10}M_\odot \left( \frac{10^{-22}~\mathrm{eV}}{m_\phi} \right)^{2/3} \left(\frac{M_{h}}{10^{12}M_\odot}\right)^{5/9},
\label{Corehalo_FDM}
\end{equation}
consistent with the expectation that the virial energy of the soliton core and that of the host halo are of the same order.
Building on this idea, Padilla et al. \cite{Padilla:2020sjy} extended the core-halo mass relation, to the case of self-interacting ULDM, showing that a similar virial energy correspondence also holds in the presence of repulsive self-interaction.
The relation in the strongly self-interacting regime is given by 
\begin{equation}
M_{\mathrm{sol}}=3.66\times 10^{10}M_\odot \left( \frac{10~\mathrm{eV}}{m_\phi / \lambda^{1/4}} \right) \left(\frac{M_{h}}{10^{12}M_\odot}\right)^{5/6}.
\label{Corehalo_SIULDM_TF}
\end{equation}
In the presence of repulsive self-interaction, the soliton becomes more massive and extended compared to the FDM case.
In this work, we use the soliton mass from Eq. \ref{Corehalo_SIULDM_TF} for Thomas-Fermi limit, fixing the total halo mass to $M_h=M_\mathrm{Fornax}$ while varying $m_\phi$ and $\lambda$.
Chavanis \cite{Chavanis_2021} showed that the Thomas-Fermi approximation becomes well valid once the particle mass exceeds 
$m_\phi = 2.92 \times 10^{-22}\,\mathrm{eV}$ and the scattering length is much larger than 
$a_s = 8.13 \times 10^{-62}\,\mathrm{fm}$, which corresponds to 
$\lambda = \frac{8\pi a_s m_\phi c}{\hbar} = 3.02 \times 10^{-90}$.
In this work, we will restrict our analysis to the regime where both the particle mass and the self-interaction strength exceed these limits.



The infall timescale due to DF in ULDM is given by~\cite{Hui:2016ltb}
\begin{equation}
\tau\equiv \frac{L}{r_0 |{\bf F}_\mathrm{DF}\cdot\hat{\bf \phi}|}=\frac{v_0^3}{4\pi\rho_cG^2 M_\mathrm{GC}\Im(I)}
\label{infall_timescale}
\end{equation}
where $L=M_\mathrm{GC}v_0 r_0$ is the angular momentum of GC of mass $M_\mathrm{GC}$ on a circular orbit of radius $r_0$ and orbital velocity $v_0$.
The maximum multipole is set to $\ell_\mathrm{max}=20$ to ensure numerical convergence of the multipole expansion in the calculation of the tangential DF coefficient $\Im(I)$.
We focus on GC3 and GC4, which have orbital radii of $0.43 \ \mathrm{kpc}$ and $0.24 \ \mathrm{kpc}$, respectively whereas the other three clusters (GC1, GC2, and GC5) orbit at radii larger than $1 \ \mathrm{kpc}$ (see Table 1 in~\cite{Cole:2012ns}).
Due to their proximity to the galactic center, these inner clusters are expected to experience stronger DF from the DM halo.
We adopt the observationally motivated range of $\rho_c$ as the background density for GC3 and GC4 as $10^{-3} \leq \rho_c/[M_\odot/\mathrm{pc}^3] \leq 10^{-1}$, inferred from stellar kinematics~\cite{Read:2018fxs, Walker:2005nt}.
This choice is justified by the fact that GC3 and GC4 reside well within the flat, core-like region of the soliton profile, where the DM density is approximately constant.
By using Eq. \ref{Corehalo_SIULDM_TF} for getting $\rho_c$ from Eq. \ref{param_TFlim},
constraint of $m_\phi$ and $\lambda$ is simply represented as
\begin{equation}
3.66 ~\mathrm{eV} \leq \frac{m_\phi}{\lambda^{1/4}} \leq 9.19 ~\mathrm{eV}.
\label{parameter_constrain}
\end{equation}
This implies that, in the strong self-interaction regime, the allowed $(m_\phi, \lambda)$ parameter space lies along contours of constant $m_\phi/\lambda^{1/4}$, as expected from the Thomas–Fermi scaling.
Interestingly, the allowed parameter range aligns well with other observational bounds, such as those inferred from the big bang nucleosynthesis ~\cite{Li:2013nal}
or from the core radii of dwarf galaxies ~\cite{Diez-Tejedor:2014naa}.

Likewise, for the FDM limit, we derive a constraint on $m_\phi$ by applying the observationally motivated $\rho_c$ range same as above to Eq. \ref{param_emp} and Eq. \ref{Corehalo_FDM}, then yields
\begin{equation}
1.17\times 10^{-22}~\mathrm{eV} \leq m_\phi \leq 4.68\times 10^{-22}~\mathrm{eV}.
\label{parameter_constrain_FDM}
\end{equation}
This range of $m_\phi$, centered around the order of $10^{-22}~\mathrm{eV}$, has been widely discussed in the context of FDM models aimed at addressing small-scale structure problems~\cite{Hui:2016ltb, Ferreira:2020fam, Hui:2021tkt}.
However, it corresponds to a relatively narrow window in parameter space and might be in tension with Lyman-$\alpha$ forest constraints, which typically require $m_\phi \gtrsim 10^{-21}~\mathrm{eV}$~\cite{Irsic:2017yje,Armengaud:2017nkf}.

\subsection{Infall timescale for leading-order approximation}\label{sec4.A}
In this subsection, we evaluate the infall timescale by replacing the tangential DF coefficient $\Im(I)$ in Eq. \ref{infall_timescale} with its leading-order approximations: $\frac{1}{3}\mathcal{M}^3=
\frac{1}{3}\left(\frac{v_0}{c_s}\right)^3$ in Eq. \ref{ImI_l=1} for the Thomas-Fermi limit and $\frac{1}{6}R_\Omega^{3/2}=\frac{\sqrt{2}}{3}\left(\frac{m_\phi v_0 r_0}{\hbar}\right)^{3/2}$ in Eq. \ref{ImI_l=1_FDM} for the FDM limit.
This approach is both analytically tractable and computational efficient, and is well-suited for the subsonic regime relevant to Fornax GCs.
The resulting expressions for the infall timescale are
\begin{equation}
\bar{\tau}=\frac{3}{4\pi\rho_cG^2 M_\mathrm{GC}}\times \begin{cases}
c_s^3 & (\mathrm{ Thomas\ Fermi \ limit}) \\
\frac{1}{\sqrt{2}} \left(\frac{\hbar v_0}{m_\phi r_0} \right)^{3/2}& (\mathrm{FDM \ limit}).
\end{cases}
\label{infall_timescale_approx}
\end{equation}
This approximation is useful for both our analytical and numerical analysis of $\tau$.
In the strongly self-interacting regime (Eq.~\ref{Corehalo_SIULDM_TF}), the relation $c_s^2 \propto \rho_c$ implies that the infall timescale scales as $\bar{\tau}\propto \sqrt{\rho_c}$ and the direct dependence on $m_\phi/\lambda^{1/4}$, allowing simultaneous constraints on ULDM particle mass and self-coupling constant from both $\rho_c$ and the leading-order approximation of DF coefficient.
We roughly set the infall timescale range as $5 ~\mathrm{Gyr} \leq \bar{\tau} \leq 20~\mathrm{Gyr}$.
The upper bound arises from the observation that some galaxies have destroyed GCs within the age of the universe ~\cite{2024A&A...683A.150M}, possibly due to DF, meaning that the upper bound is not very stringent.
Using the masses and radial distances of GC3 and GC4 in columns 1 \& 2 in Table \ref{Timescale_Table}, respectively, and Eq. \ref{infall_timescale_approx}, the results are 
\begin{equation}
3.75 ~\mathrm{eV} \leq \frac{m_\phi}{\lambda^{1/4}}\leq 5.57 ~\mathrm{eV} \ (\mathrm{GC3}) \ \ \ , \ \ \ 5.00 ~\mathrm{eV} \leq \frac{m_\phi}{\lambda^{1/4}} \leq 7.44 ~\mathrm{eV} \ (\mathrm{GC4}),
\label{parameter_constrain_from_tau}
\end{equation}
which are consistent with other cosmological bounds ~\cite{Dave:2023wjq,Li:2013nal,Diez-Tejedor:2014naa,Lee:2024pxl}.

\begin{figure*}[!ht]
\centering
\includegraphics[width=0.9\textwidth]{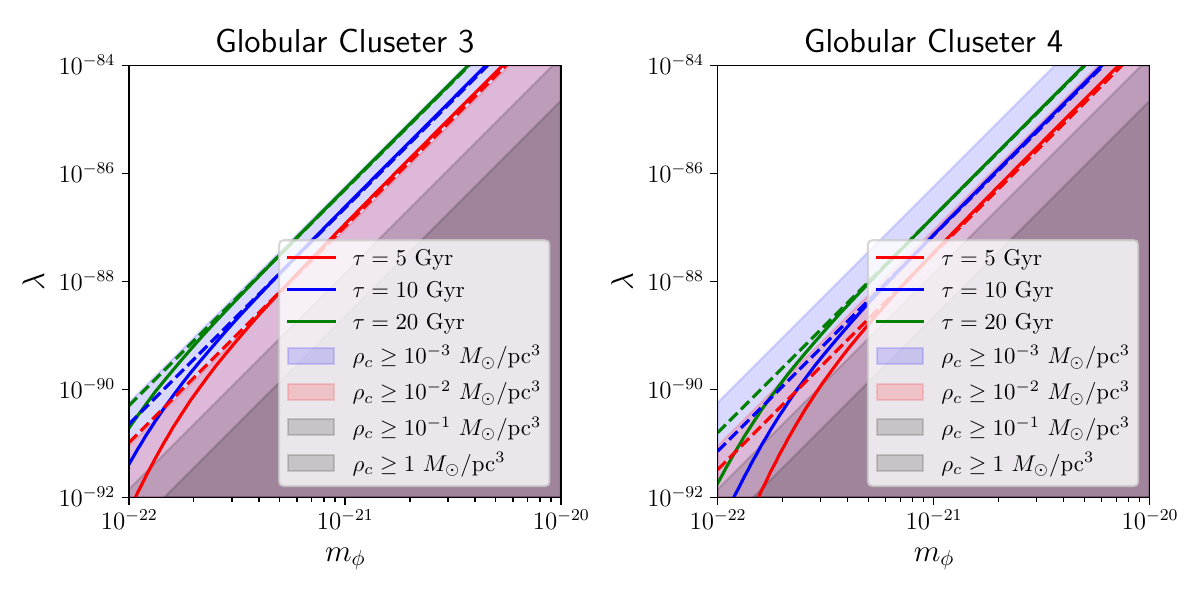}
 \caption{
 Contours of infall timescales 5, 10, 20~Gyr for full-order (solid) and leading-order (dashed) calculations are plotted in the $(m_\phi,~\lambda)$ parameter space for GC3 (left) and GC4 (right). 
 The shaded regions correspond to the central density of Fornax dSph, satisfying $\rho_c \geq 10^{-3},~10^{-2},~10^{-1},~1 \ M_\odot / \mathrm{pc}^3$.
 }
 \label{Parameter_Lead}
\end{figure*}

\begin{figure*}[!ht]
\centering
\includegraphics[width=0.55\textwidth]{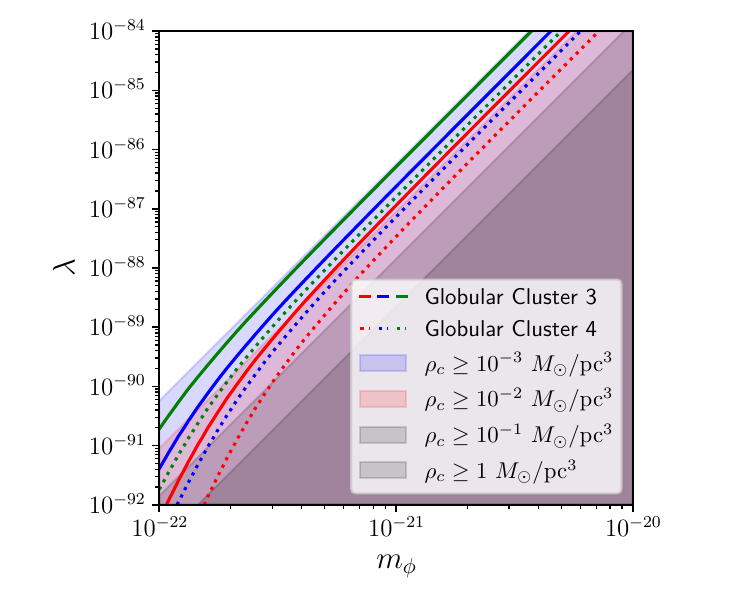}
 \caption{
  Direct comparison between the full-order infall timescales of GC3 (solid) and GC4 (dotted).
 }
 \label{Parameter_Mix}
\end{figure*}

Fig. \ref{Parameter_Lead} visualizes the infall timescales 5, 10, 20~Gyr in the $(m_\phi,~\lambda)$ parameter space for both full-order ($\tau$) and leading-order ($\bar{\tau}$) calculations.
These represent characteristic timescales comparable to or exceeding the observed lifetimes of the Fornax GCs.
We also overlay regions of soliton central density $\rho_c \ge 10^{-3}, ~10^{-2}, ~10^{-1}$ and $1\ M_\odot /\mathrm{pc}^3$, based on observational estimates of Fornax dSph.
Their boundaries are approximately parallel to $\bar{\tau}$ contours for $\lambda \gtrsim 10^{-88}$ roughly, reflecting their shared dependence on $\rho_c$ in the Thomas-Fermi limit.
At larger $\lambda$ or $m_\phi$, the $\tau$ contours increasingly align with lines of constant $m_\phi/\lambda^{1/4}$, consistent with the Thomas-Fermi limit.
Correspondingly, the full-order and leading-order results closely agree in this regime, indicating that higher-order corrections are negligible.
By contrast, this scaling behavior breaks down for smaller values of $\lambda$ or $m_\phi$ (roughly $\lambda \lesssim 10^{-90}$), where the self-interaction is no longer sufficient to sustain the Thomas-Fermi limit.
In this regime, quantum pressure dominates, causing deviations from the linear $m_\phi/\lambda^{1/4}$ scaling and from the leading-order prediction (details in Sect.II.G of~\cite{Chavanis:2011zi} and Appendix \ref{App_Validity}).
A comparison of full-order infall timescales between GC3 and GC4 is shown in Fig. \ref{Parameter_Mix}, which enables a direct assessment of their relative dynamical behavior across the $(m_\phi,~\lambda)$ parameter space.

Similarly, we calculate $\bar{\tau}$ for FDM limit, using the second expression in Eq. \ref{infall_timescale_approx}.
Due to the assumption that the background density is set to $\rho_c$, the orbital velocity of each GC is given by $v_0 = \sqrt{\frac{GM_\mathrm{in}(r_0)}{r_0}}$ where the ULDM mass enclosed within the orbit of GC is $M_\mathrm{in}(r_0)=\frac{4}{3}\pi r_0^3 \rho_c$.
The results are
\begin{equation}
1.01\times 10^{-22} ~\mathrm{eV} \leq m_\phi\leq 1.83\times 10^{-22} ~\mathrm{eV} \ (\mathrm{GC3}) \ \ \ , \ \ \ 1.56 \times 10^{-22}~\mathrm{eV} \leq m_\phi \leq 2.83 \times 10^{-22}~\mathrm{eV} \ (\mathrm{GC4}).
\label{parameter_constrain_FDM_from_tau}
\end{equation}
The above ranges of $m_\phi$ partially overlap with those derived from Eq. \ref{parameter_constrain_FDM}, but they correspond to a much narrower parameter space.
This is significant because, for FDM, some observational data ~\cite{Irsic:2017yje,Armengaud:2017nkf,Zimmermann:2024xvd} suggest constraints on $m_\phi \gtrsim 10^{-21}~\text{eV}$,
which could mean that FDM might not be a viable solution to the timing problem
for $m_\phi \gtrsim 10^{-21}~\text{eV}$.
For instance, when adopting $m_\phi=10^{-21}~\mathrm{eV}$, the resulting infall timescales are merely $\sim 0.096~\mathrm{Gyr}$ for GC3 and $\sim 0.263~\mathrm{Gyr}$ for GC4, which further underscores that the timing problem cannot be resolved within the mass range favored by Lyman-$\alpha$ forest constraints.
This bound is also noted in the DF calculation for a linear motion~\cite{Lancaster:2019mde}.

Table \ref{Timescale_Table} summarizes the full-order ($\tau$) and leading-order ($\bar{\tau}$) infall timescales for GC3 and GC4 in both the FDM and self-interacting ULDM models, for $m_{\phi,1} = 5\times 10^{-22}~\mathrm{eV}$ and $m_{\phi,2} = 10^{-21}~\mathrm{eV}$. 
The $m_{\phi,2}$ case is included to probe the regime relevant to Lyman-$\alpha$ constraints and Thomas–Fermi limit, while its FDM results are omitted due to the extremely short timescales noted earlier.
For the self-interacting cases, infall timescales are presented as ranges reflecting the soliton central density range $10^{-3} \leq \rho_c/[M_\odot/\mathrm{pc}^{3}] \leq 10^{-1}$.
In the Thomas–Fermi limit, this translates into the following allowed self-coupling constants: $8.69\times10^{-90} \leq \lambda \leq 3.39\times10^{-88}$ for $m_{\phi,1}$ and $1.21\times10^{-88} \leq \lambda \leq 5.18\times10^{-87}$ for $m_{\phi,2}$.
In general, the inclusion of self-interaction results in systematically longer infall timescales than those obtained in the FDM case, reflecting the suppression of DF by enhanced effective pressure.
Specifically, the minimum self-coupling constants at which the infall timescale exceeds $10~\mathrm{Gyr}$ are shown in Table \ref{lambda10Gyr_Table}, corresponding to the blue contours in Figs. \ref{Parameter_Lead} and \ref{Parameter_Mix}.
This indicates that the leading-order approximation provides a reliable estimate in regimes capable of resolving the timing problem of Fornax GCs.
These results demonstrate that sufficient self-interaction is essential for reproducing the observed lifetimes of the Fornax GCs, whereas the FDM limit fails to do so within the mass range allowed by other astrophysical constraints.

\begin{table}[!ht]
\centering
\caption{
Columns 1 and 2 present the mass $M_\mathrm{GC}$ and the projected radius $r_0$ of the Fornax globular clusters GC3 and GC4, which are taken from~\cite{Cole:2012ns}.
Columns 3$\sim$6 present the infall timescales for $m_{\phi,1} = 5\times 10^{-22}~\mathrm{eV}$, calculated in the FDM limit (cols. 3--4) and considering self-interaction (cols. 5--6) using full-order and leading-order DF coefficients, respectively.
Columns 7 and 8 show the corresponding results only for self-interacting ULDM with $m_{\phi,2} = 10^{-21}~\mathrm{eV}$.
All calculations for columns 5$\sim$8 reflect not only the range of soliton central density $10^{-3} \leq \rho_c/[M_\odot/\mathrm{pc}^3] \leq 10^{-1}$, but also the range of self-coupling constants: $8.69\times10^{-90} \leq \lambda \leq 3.39\times10^{-88}$ for $m_{\phi,1}$ and $1.21\times10^{-88} \leq \lambda \leq 5.18\times10^{-87}$ for $m_{\phi,2}$.
These results highlight the role of self-interaction in suppressing DF, whereas for the FDM case with $m_{\phi,1}$ is in tension with some observational constraints, such as from the Lyman-$\alpha$ forest.
}
\begin{tabular}[t]{c||c|c|c|c|c|c|c|c}
\hline
& \multirow{4}{*}{$M_\mathrm{GC}$ [$ M_\odot$]}&\multirow{4}{*}{$r_0$ [kpc]} & \multicolumn{6}{c}{$\tau$, $\bar{\tau}$ [Gyr]} 
\\ \cline{4-9} & & & \multicolumn{4}{c|}{$m_{\phi,1}=5\times 10^{-22}~\mathrm{eV}$} & \multicolumn{2}{c}{$m_{\phi,2}= 10^{-21}~\mathrm{eV}$} \\
\cline{4-9}
& & & \multicolumn{2}{c|}{FDM} & \multicolumn{2}{c|}{SI-ULDM} & \multicolumn{2}{c}{SI-ULDM} \\
\cline{4-9}
& & & Full Order & Leading Order & Full Order & Leading Order & Full Order & Leading Order \\
\hline\hline

GC3&$3.63\times 10^5$&0.43&0.71&0.48&1.07$~\sim~$21.4&0.86$~\sim~$21.2&0.54$~\sim~$20.0&0.76$~\sim~$20.4\\
GC4&$1.32\times 10^5$&0.24&1.46&1.33&3.20$~\sim~$59.7&2.36$~\sim~$58.4&1.95$~\sim~$56.0&2.09$~\sim~$56.1\\
\hline
\end{tabular}
\label{Timescale_Table}
\end{table}

\begin{table}[!ht]
\centering
\caption{
Minimum self-coupling constants ($\lambda$) of SI-ULDM for which the infall timescales of Fornax GCs exceed $10~\mathrm{Gyr}$, obtained from both full-order and leading-order DF coefficients for $m_{\phi,1}$ (cols. 1--2) and $m_{\phi,2}$ (cols. 3--4). 
These correspond to the blue contours in Figs. \ref{Parameter_Lead} and \ref{Parameter_Mix} and mark the $(m_\phi,~\lambda)$ parameter space capable of resolving the timing problem of the Fornax GCs.
}
\begin{tabular}[t]{c||c|c|c|c}
\hline
& \multicolumn{2}{c|}{$m_{\phi,1}=5\times 10^{-22}~\mathrm{eV}$} & \multicolumn{2}{c}{$m_{\phi,2}= 10^{-21}~\mathrm{eV}$}
\\ \cline{2-5} & Full Order & Leading Order & Full Order & Leading Order\\
\hline\hline 
GC3 & $1.33\times10^{-88}$&$1.35\times10^{-88}$&$2.44\times10^{-87}$&$2.44\times10^{-87}$ \\
GC4 & $3.91\times10^{-89}$&$4.25\times10^{-89}$&$ 7.91\times10^{-88}$&$7.91\times10^{-88}$ \\
\hline
\end{tabular}
\label{lambda10Gyr_Table}
\end{table}

\section{Summary}\label{sec5}
In this paper, we studied the impact of repulsive self-interaction in ULDM on dynamical friction (DF) acting on circular orbits, with a focus on the timing problem of Fornax dwarf spheroidal (dSph) galaxy's globular clusters (GCs).
We employed the Gross-Pitaevskii-Poisson system, which effectively describes ULDM in the non-relativistic regime.
Based on this framework, we first examined two soliton density profiles to describe non-interacting and strongly self-interacting regimes of ULDM halo.
We then derived the DF force acting on a steady circular orbit by applying linear perturbation theory within the Madelung formalism.
The DF coefficient is encoded in a complex parameter $I$, whose real and imaginary parts correspond to the radial and tangential DF coefficients, and can be represented via multipole expansion.
We also derived the leading-order behavior of $I$ as a function of the Mach number $\mathcal{M}$, defined as the ratio of orbital velocity to the sound speed induced by the ULDM self-interaction.
In the FDM limit, the leading-order contribution was evaluated in terms of the characteristic angular momentum $R_\Omega$, which encodes the orbital scale relative to the de Broglie wavelength.
It has been established that the DF in FDM is weaker than in CDM, and our results confirm that including self-interaction further suppresses the DF effect, and avoids the tension of ULDM particle mass from several astronomical observations.
Considering only the leading-order term simplifies the calculation in subsonic motion, where the tangential DF force dominates.
These leading-order approximations capture the essential dynamics of the Fornax GCs, as supported by orbiting velocity estimates derived from radial velocity observations.

The DF results were applied to calculate the infall timescales of two Fornax globular clusters, GC3 and GC4.
As a result, the infall timescales were found to increase with strong self-interaction, allowing the timing problem of the Fornax dwarf spheroidal (dSph) to be addressed more effectively than in the case of FDM.
We calculated the infall timescales using only the leading-order term of the tangential DF coefficient, and found that it yields better agreement with the full-order results for stronger self-interaction.
Furthermore, by assigning specific ranges to the infall timescales of each globular cluster, the parameter $m_\phi/\lambda^{1/4}$ can be constrained accordingly.
This provides a broader parameter space in which self-interacting ULDM can be tested for consistency with cosmological and other astrophysical observations. 
These results offer new insights into the dynamics of dwarf galaxies and highlight the potential of self-interacting ULDM to address small-scale challenges in DM physics.

Despite the extended infall timescales achievable through self-interaction, it remains to be seen whether this model avoids all other astrophysical constraints.
Although it is known that a repulsive self-interaction characterized by $m_\phi/\lambda^{1/4} \simeq 10~\mathrm{eV}$ can satisfy some cosmological constraints,
including those related to galaxy core sizes and dark matter
density ~\cite{Dave:2023wjq,Li:2013nal,Lee:2024pxl}, it remains unclear whether such an interaction can also account for the Lyman-$\alpha$ observations.
(The strong limits set by Lyman-$\alpha$ observations
for FDM may be softened once the impact of attractive self-interactions is considered ~\cite{Ferreira:2020fam,Tremaine_nuclei1}.)
Therefore, while the repulsive self-interaction offers a potential resolution to the Fornax timing problem, its full consistency with Lyman-$\alpha$ constraints and other astrophysical bounds requires further investigation. 
In this regard, the continued development of diverse numerical methods such as N-body simulations \cite{Nori:2018hud, Hopkins:2018tht}, Boltzmann solvers \cite{Hlozek:2014lca, Cembranos:2018ulm, Poulin:2018dzj,Aboubrahim:2024spa}, and grid-based approaches \cite{Li:2018kyk, Schwabe:2020eac}--for various ULDM models is expected to enable more direct and robust constraints from astronomical observations.
\begin{acknowledgments}
The authors thank Inkyu Park and Dongsu Bak for their helpful comments.
HK was supported by Basic Science Research Program through the National Research Foundation (NRF) funded by the Ministry of Education (2018R1A6A1A06024977).
\end{acknowledgments}

\begin{appendix}
\section{Validity of the Thomas-Fermi limit through dimensional analysis}\label{App_Validity}
In this appendix, we review the classification of the Thomas-Fermi and FDM limit of ULDM based on dimensional analysis~\cite{Chavanis:2011zi}, and briefly discuss these validity condition.
The steady state soliton, characterized by mass $M_\mathrm{sol}$ and size $R$, is described by setting $\frac{\partial}{\partial t}=0$ and ${\bf v}=0$ in the Euler equation (Eq. \ref{Euler_eq}).
Combining this with Poisson equation (Eq. \ref{GPP_Peq}), we obtain the condition for hydrostatic equilibrium including both quantum and barotropic pressures:
\begin{equation}
4\pi G\rho  -\frac{\hbar^2}{2m_\phi^2}\nabla^2 \frac{\nabla^2 \sqrt{\rho}}{\sqrt{\rho}}+ \frac{\lambda \hbar^3}{2m_\phi^4 c}\nabla^2 \rho =0.
\label{Hydrostatic_Equilibrium}
\end{equation}
Based on Eq. \ref{Hydrostatic_Equilibrium}, we estimate the soliton size for a density $\rho\sim M_\mathrm{sol}/R^3$ by applying dimensional analysis in two distinct limits.
In the FDM limit, where the quantum pressure is solely balancing with the gravitational collapse, we neglect the third term in Eq. \ref{Hydrostatic_Equilibrium}, leading to a characteristic size $R_Q$, often referred to as the gravitational Bohr radius~\cite{Lee:2021bje, Lee:2023krm}
\begin{equation}
R_Q=\frac{\hbar^2}{G M_\mathrm{sol} m_\phi^2}.
\label{R_Q}
\end{equation}
In the Thomas-Fermi limit, where the barotropic pressure is solely balancing with the gravitational collapse, we neglect the second term in Eq. \ref{Hydrostatic_Equilibrium}, leading to obtain $R_\lambda$ as the characteristic size scale:
\begin{equation}
R_\lambda = \sqrt{\frac{\lambda\hbar^3}{Gm_\phi^4 c}}.
\label{R_lambda}
\end{equation}
This is consistent with the soliton size $R_\mathrm{TF}$ in Eq. \ref{param_TFlim}.
Comparing these two characteristic scales, the Thomas-Fermi limit is valid when $R_\lambda \gg R_Q$~\cite{Chavanis:2011zi}.
Equivalently, the self-coupling constant must satisfy
\begin{equation}
\lambda \gg 1.198\times 10^{-92}\left(\frac{10^8 M_\odot}{M_\mathrm{sol}}\right)^2.
\label{lambda_TFValidRegime}
\end{equation}
Conversely, the limit $\lambda\rightarrow 0$, where $R_\lambda \ll R_Q$, indicates the breakdown of the Thomas-Fermi limit.
In this regime, the ULDM system reduces to the FDM limit, where it is well described by the Schr\"odinger-Poisson equations.
As $\lambda$ increases, the associated sound speed $c_s$ in Eq. \ref{sound_speed} becomes larger, resulting in a lower Mach number $\mathcal{M}$. 
This enhances the validity of the low-Mach number approximation, particularly for the leading-order term of the tangential DF coefficient $\Im(I)$.

\section{Evaluating leading-order term in $S_{\ell,\ell-1}^m$} \label{App_sphericalbessel}
In this appendix, we compute the leading-order term of the imaginary part of the DF coefficient $I$ in its multipole expansion, represented in Eq. \ref{Result_I}.
For $\ell=1$, where the leading-order term only appears, we could focus on the term
\begin{equation}
\Im(I)_{\ell=1}=-\frac{2}{\pi}\Im\left(S_{1,0}^{-1}-S_{1,0}^0\right).
\end{equation}

Using the standard definitions of the modified spherical Bessel functions: $i_\ell (x) \equiv i^{-\ell} j_\ell (ix)$ and $k_{\ell-1}(x) \equiv -i^{\ell-1}h_{\ell-1}^{(1)}(ix)$, we obtain
\begin{equation}
j_{\ell}(ix)h_{\ell-1}^{(1)}(ix)=(-i)i_\ell (x) k_{\ell-1} (x),
\end{equation}
which is a purely imaginary function.
This identity generally simplifies the calculation of $S_{\ell,\ell-1}^m$ both in the self-interacting (Eq. \ref{Result_S}) and FDM (Eq. \ref{Result_S_FDM}) cases.
Thus, this identity directly guarantees that $S_{\ell,\ell-1}^0$ is real for all $\ell$ in the self-interacting regime.
The case for the FDM limit is somewhat more subtle due to the infrared divergence in the calculation of the $S_{1,0}^0$ integral; however, its contribution remains purely real~\cite{DFrot_FDM, Berezhiani:2023vlo}.

For $S_{1,0}^{-1}$, both the self-interacting and FDM limit take the common form $\sim (-i)[j_1(x_1)h_0^{(2)}(x_1)+j_1(ix_2)h_0^{(1)}(ix_2)]$.
Using the asymptotic expansions valid in the limit $x_1,x_2 \ll 1$, the leading-order term is $\sim \frac{1}{3}(x_2 - i x_1)$, leading to
\begin{equation}
\Im \left\{(-i)\left[j_1(x_1)h_0^{(2)}(x_1)+j_1(ix_2)h_0^{(1)}(ix_2)\right] \right\}= -\frac{1}{3}x_1 .
\end{equation}
For $\ell_q \gg 1$, the substitutions $x_1\rightarrow \ell_q \mathcal{M}f_{-1}^- \simeq \mathcal{M}$ and $x_2\rightarrow \ell_q \mathcal{M}f_{-1}^+ \simeq 2m_\phi c_s r_0 / \hbar$, followed by their insertion into Eq. \ref{Result_S}, yield the leading-order approximation of the tangential DF coefficient in the strongly self-interacting regime, as shown in Eq. \ref{ImI_l=1}.
Similarly, substituting $x_1=x_2 \rightarrow \sqrt{R_\Omega}$ into Eq. \ref{Result_S_FDM} yields the leading-order approximation of the tangential DF coefficient in the FDM limit, as presented in Eq. \ref{ImI_l=1_FDM}.

\end{appendix}

\bibliography{main}{}

\end{document}